\documentclass[%
 reprint,
superscriptaddress,
 amsmath,amssymb,
 aps,
prb,
]{revtex4-1}

\usepackage{graphicx}
\usepackage{dcolumn}
\usepackage{bm}
\usepackage{cleveref}

\newcommand{\BaFeCoAs}{Ba(Fe$_{1-x}$Co$_{x}$)$_{2}$As$_{2}$ }
\newcommand{\BaFeCoAsNS}{Ba(Fe$_{1-x}$Co$_{x}$)$_{2}$As$_{2}$}
\begin{document}

\preprint{APS/123-QED}

\title{Signatures of Anelastic Domain Relaxation in \BaFeCoAs Investigated by Mechanical Modulation of Resistivity}

\author{Alexander T. Hristov}
\affiliation{Geballe Laboratory for Advanced Materials, Stanford University, Stanford, CA 94305}
\affiliation{Stanford Institute for Materials and Energy Science, SLAC National Accelerator Laboratory, 2575  Sand  Hill  Road,  Menlo  Park,  California  94025,  USA}
\affiliation{Department of Physics, Stanford University, Stanford, CA 94305}

\author{Matthias S. Ikeda}
\affiliation{Geballe Laboratory for Advanced Materials, Stanford University, Stanford, CA 94305}
\affiliation{Stanford Institute for Materials and Energy Science, SLAC National Accelerator Laboratory, 2575  Sand  Hill  Road,  Menlo  Park,  California  94025,  USA}
\affiliation{Department of Applied Physics, Stanford University, Stanford, CA 94305}

\author{Johanna C. Palmstrom}
\affiliation{Geballe Laboratory for Advanced Materials, Stanford University, Stanford, CA 94305}
\affiliation{Stanford Institute for Materials and Energy Science, SLAC National Accelerator Laboratory, 2575  Sand  Hill  Road,  Menlo  Park,  California  94025,  USA}
\affiliation{Department of Applied Physics, Stanford University, Stanford, CA 94305}

\author{Ian R. Fisher}
\affiliation{Geballe Laboratory for Advanced Materials, Stanford University, Stanford, CA 94305}
\affiliation{Stanford Institute for Materials and Energy Science, SLAC National Accelerator Laboratory, 2575  Sand  Hill  Road,  Menlo  Park,  California  94025,  USA}
\affiliation{Department of Applied Physics, Stanford University, Stanford, CA 94305}

\date{\today}

\begin{abstract}
The resistive response of \BaFeCoAs to AC mechanical deformation is considered in the multi-domain state.
This resistance change to external forces depends both upon the anelastic relaxation of domain walls and upon the relation between resistance and the domain wall configuration.
In this experiment, samples are adhered to the surface of a piezoelectric stack, which is driven by an AC voltage while the AC modulation of the sample resistance is measured.
As the response time of electrons is faster than that of the lattice, the phase difference $\phi$ between the AC resistance modulation and the AC deformation of the piezoelectric is entirely due to anelastic relaxation effects in the sample.
An expression is derived for relating $\phi$, the phase lag between the resistance of the sample and the deformation of the driving piezoelectric, to a sample's complex compliance, $J(\omega)$, in this experimental configuration.
Measurements of \BaFeCoAs for $x$= (0.025, 0.052) reveal a peak in the out-of-phase resistivity modulation in the orthorhombic antiferromagnetic state well below the Ne\'el temperature $T_N$ and structural transition $T_S$.
Meanwhile, for a composition that is tetragonal at all temperatures, $x$= 0.07, the resistance modulation remains entirely in phase over the same temperatures, establishing domain motion as a probable cause of the observed effects in the samples that do undergo the tetragonal-to-orthorhombic phase transition.
Fits are provided of $\tan\phi$ for a sample with $x=0.025$ for various amplitude excitations on the piezoelectric stack, from which the apparent activation energy $E_a$ for domain wall motion is found.
In these measurements, $E_a$ decreases with increasing amplitude of the deformation along the $[110]_T$ axis. 
Measuring the amplitude of deformation from the deformation of the piezoelectric stack aligned to this axis, $\varepsilon^0_{[110,110]_T}$, we find $\frac{dE_a}{d\varepsilon^0_{[110,110]_T}} = -1115\pm 196$ eV 
with $E_a = 9.09 \pm 0.74 \times 10^{-3}$ eV in the zero strain limit if we assume linearity over the entire strain range.
\end{abstract}

\maketitle


\section{Introduction}

In the theory of anelastic relaxation, the mechanical response of a material to a sudden, externally induced stress is decomposed into an instantaneous elastic component and a slowly-relaxing anelastic component.
The latter arises when an internal degree of freedom simultaneously couples to the lattice and has dynamics governed by a slow relaxation mechanism. \cite{AnelasticBook}
For example, when an external force is applied to a sample with local anistropic defects, anelastic relaxation will be observed because the anisotropic defects inside the sample only relax slowly via a thermally activated tunneling processes towards a low energy state in which they are aligned with the stress-induced lattice deformation.

Anelastic relaxation is typically manifested in mechanical measurements, such as those of the strain, $\varepsilon$, induced by an external stress, $\sigma$, (known as the J-type response function), or vice versa (known as the E-type response). 
When the relaxation rate of the internal degree of freedom is uniform and proportional to its distance from equilibrium, the result is an exponential relaxation process with a single relaxation rate $\tau$ that specifies a Debye spectrum for the response functions (i.e. the compliance tensor),
\begin{equation}
J(\omega) = \frac{\varepsilon (\omega)}{\sigma(\omega)} = J_u + (J_r - J_u) \frac{ 1}{ 1 + i \omega \tau}
\label{eq:responsefunctiondef}
\end{equation}
where $J_u$ is known as the un-relaxed ($\omega \to \infty$ or $t \to 0$) response and $J_r$ is the relaxed ($\omega \to 0$ or $t \to \infty$) response.
As the functional form clearly indicates $J(\omega) = J_1(\omega) - i J_2(\omega) $ is a complex quantity with real and imaginary components given by $J_1$ and $- i J_2(\omega)$ respectively.
For measurements of the phase difference between the stress and strain, 
\begin{equation}
\tan(\phi)=\frac{J_2(\omega)}{J_1(\omega)} \approx \frac{J_r-J_u}{J_u}\frac{\omega\tau} {1 + \omega^2\tau^2}
\label{eq:tan}
\end{equation}
provided that the common condition ${J_r-J_u}<<{J_u}$ is satisfied.

For the case where the relaxation process is the result of thermal activation the relaxation time depends on temperature according to the Arrhenius law 
\begin{equation}
\tau = \tau_0 e^{E_a/k_bT},
\label{eq:arrhenius}
\end{equation} 
for an activation barrier $E_a$, while for cooperative freezing transitions at a temperature $T_f$, the empirical Vogel-Fulcher law, 
\begin{equation}
\tau = \tau_0 e^{E_a/k_b(T-T_f)},
\label{eq:vogel}
\end{equation} 
is widely used.
Interactions between internal degrees of freedom, or multiple inhomogeneous relaxation rates, as detailed in ref. \onlinecite{NowickBerryIBM} and later in this document, can make important changes to the microscopic parameters inferred from fits of $J(\omega,T)$. 
However, it is also worth noting such changes typically have little effect on the qualitative form of the response function other than by widening the temperature and frequency scale over which features in the response function emerge, or making such features slightly more asymmetric with respect to a tuning parameter like frequency or temperature.

Measurements of the dynamical response functions can be obtained from direct measurement of the stress and strain, as well as by mechanical resonance techniques\cite{AnelasticBook}.
In this work, we demonstrate how the dynamical response function can also be measured through the electronic rather than mechanical response of the material.
Electronic degrees of freedom relax much faster than the crystal lattice, and so electronic properties like the resistance provide a near instantaneous measure of domain reconfiguration and the associated changes in the strain $\varepsilon$ in a sample.
In a number of materials with compound or multi-ferroic order, orbital and magnetic degrees of freedom are intertwined with the lattice and necessarily respond to lattice relaxation processes, with the additional benefit of being addressable by high precision electrical, optical or magnetic measurement techniques.

The \BaFeCoAs alloy series has a rich phase diagram featuring superconductivity, antiferromagnetism and electronic-nematic order\cite{Chu_Determination_2009,Ni2008}. 
While the high temperature crystal structure is tetragonal, successive electronic-nematic and antiferromagnetic transitions occur upon cooling in the regime $0\leq x \leq 0.06$. 
The former of these breaks the mirror planes normal to the $[010]$ and $[100]$ tetragonal crystallographic axes, which manifests in several ways, including a large resistivity anisotropy between the orthorhombic a and b axes (which are aligned to the tetragonal $[110]$ and $[1\bar{1}0]$ axes)\cite{Chu824,TanatarPRL}.
Applying stress along these axes induces reconfiguration of domains in the sample, which change both the resistivity and current paths inside the sample, and manifests as a change of the sample resistance.
We therefore measure the modulation of the resistance of samples of \BaFeCoAs in the multi-domain state by an external AC mechanical manipulation provided by a piezoelectric device.
Using this approach we identify clear signatures of anelastic relaxation in measurements well below the nematic and antiferromagnetic transitions at frequencies ranging from 1~Hz to 3.3~kHz, and we estimate the characteristic energy scales associated with domain wall pinning in these materials.

\section{Mechanical Deformation Of An Anelastic Sample on a Piezoelectric Stack}

Previous resistance measurements of samples under both quasi-static and dynamical mechanical deformation have demonstrated that the mechanical deformation of the sample and piezoelectric stack track one another closely in both amplitude and phase in the tetragonal state, where the system has no orthorhombic domains.
Consequently, one can measure the approximate displacement of mono-domain crystals by affixing a strain gauge on to the opposite surface of the piezoelectric stack.\cite{HristovMeasurement2018}

\begin{figure}
\includegraphics[width=\linewidth]{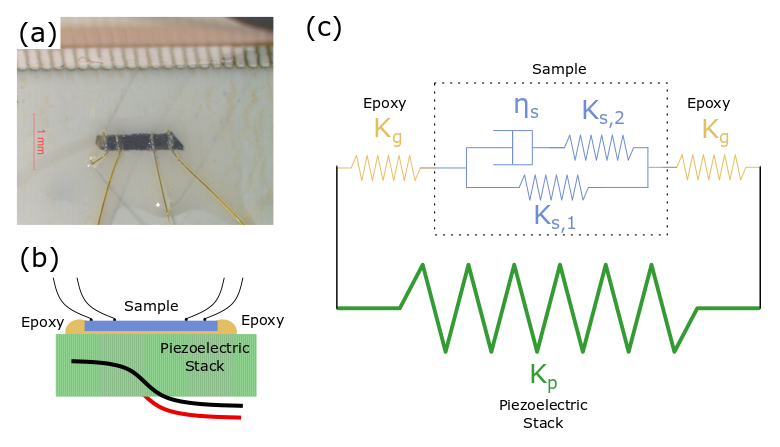}
\caption{
(a) Top view of a sample of Ba(Fe$_{1-x}$Co$_{x}$)$_{2}$As$_{2}$ (x=0.052), measuring approximately 1580$\mu$m $\times$ 260 $\mu$m, with an approximate thickness of $0.02$ to $0.04$~mm, that was attached to the surface of a piezoelectric stack for experiments. 
(b) Schematic side view of the same sample.  
(c) Physical model of the sample (contents of the dashed box), adhesive on the ends of the sample (springs annotated $K_g$) and the larger piezoelectric stack (spring annotated $K_p$). The colors are matched between panels (b) and (c) to highlight the physical correspondence. The excitation voltage changes the unstressed length of the piezoelectric stack, which drives mechanical changes through the rest of the system. This model is solved in the text.
\label{fig:schematic}
}
\end{figure}

A sample displaying clear signatures of anelastic relaxation, whether due to domain wall motion or other effects, necessarily deviates from exactly tracking the piezoelectric stack.
Therefore a new model must be developed for understanding how the displacement of the piezoelectric stack drives mechanical properties.

\subsection{Debye Standard Linear Solid on a Piezoelectric Stack}

For a start, we refer readers to \cref{fig:schematic}, which holds all the essential elements for studying an anelastic sample adhered to a piezoelectric stack, with the relevant parameters {$K_i$}, $\eta$ suitably defined.   
The sample, consisting of elements inside the outlined box, is treated as a standard linear solid, for which the stress-strain relation is given by
\begin{equation}
\sigma +\frac{\eta}{K_{s,2}}\dot{\sigma} = K_{s,1} \varepsilon + \frac{\eta_s}{K_{s,2}} \left(K_{s,1} + K_{s,2} \right) \dot{\varepsilon}. 
\label{eq:standardlinearsolid}
\end{equation}
For a sample affixed to a piezoelectric stack, neither the stress, nor the strain, is immediately easy to measure.
Instead, we rely on the model displayed in \cref{fig:schematic} to relate the stress to other experimentally accessible quantities, and we use the framework below to explain how the mechanical strain configuration of the sample can be tracked by the sample resistance.

The piezoelectric stack is treated as a simple spring with an unstressed length that depends on voltage, such that a voltage oscillation on the piezoelectric stack will drive deformation of all components of this model.
It is known that these stacks are hysteretic, and so the mechanical deformation of the piezoelectric stack was monitored by attaching a WK-05-062TT-350L (Vishay Precision Group) resistive strain gauge to the opposite surface of the piezoelectric stack, so that $\varepsilon_p$ is known.
Furthermore, the two elements corresponding to the glue are taken to be symmetric for computational simplicity.

To solve this model, the net stress on the end points must be zero, and the glue and sample are in series, so it must be that 
\begin{equation}
-\sigma_p =\sigma_g =\sigma_s \equiv \sigma
\label{eq:sigmas}
\end{equation}
which represents the stress on the piezoelectric, glue and sample, respectively.
Furthermore, 
\begin{equation}
L_{0p}\varepsilon_p = 2L_{0g} \varepsilon_g + L_{0s}\varepsilon_s
\label{eq:lengths}
\end{equation}
where $L_{0p}, L_{0g}$ and $L_{0s}$ are the active (meaning only the parts of the piezoelectric between the boundaries of the glue are counted) lengths of the piezoelectric, glue and sample when the sample is unstrained. 
Combining \cref{eq:sigmas} and \cref{eq:lengths} with the property that $\sigma = K_g \varepsilon_g$ for the glue in this model, we obtain
\begin{equation}
\sigma = K_g \frac{L_{0p}\varepsilon_p - L_{0s}\varepsilon_s}{2L_{0g} } 
\label{eq:samplesigma}
\end{equation}
which is then substituted into \cref{eq:standardlinearsolid} to obtain the relation between the piezoelectric stack and sample
\begin{equation}
\Gamma = \Gamma_1 -i \Gamma_2 =\frac{\varepsilon_s}{\varepsilon_p} = \frac{\left[1 +\omega^2 \tau^2 (1+\Delta)\right] - i \omega \tau \Delta}{1 +\omega^2\tau^2(1+\Delta)^2}
\label{eq:gamma}
\end{equation}
where we have used $\Delta = \frac{K_{s,2}}{K_{s,1} + K_g (\frac{L_{0s}}{2 L_{0g}})}$ and $\tau = \eta_s / K_{s,2}$.
Then, the phase difference between the strain on the piezoelectric stack and the mechanical state of the sample is
\begin{equation}
\tan\phi = \frac{\Gamma_2}{\Gamma_1} = \frac{\Delta \omega\tau}{1 + \omega^2 \tau^2 (1 + \Delta)} = \frac{\tilde{\Delta}\omega\tilde{\tau}}{1+\omega^2\tilde{\tau}^2},
\label{eq:tanphi2}
\end{equation}
for $\tilde{\tau} = \tau\sqrt{1+\Delta}$ and $\tilde{\Delta} = \Delta /\sqrt{1+\Delta}$.

Importantly, the expression in \cref{eq:tanphi2} has an identical functional form to that of \cref{eq:tan}, with a simple mapping between ($\tilde{\tau}$, $\tilde{\Delta}$) and ($\tau$, $\Delta$). 
Furthermore, we note that a relaxation time given by either Arrhenius or Vogel-Fulcher behavior, the effect of this re-scaling is only in $\tau_0$ and not the activation potential $E_a$.
Conceptually, this result is not a surprise, as the glue and piezoelectric are treated as predominantly elastic components, so that their displacement and the global stress field are in phase.

\begin{figure*}
    \centering
    \includegraphics[width =\textwidth]{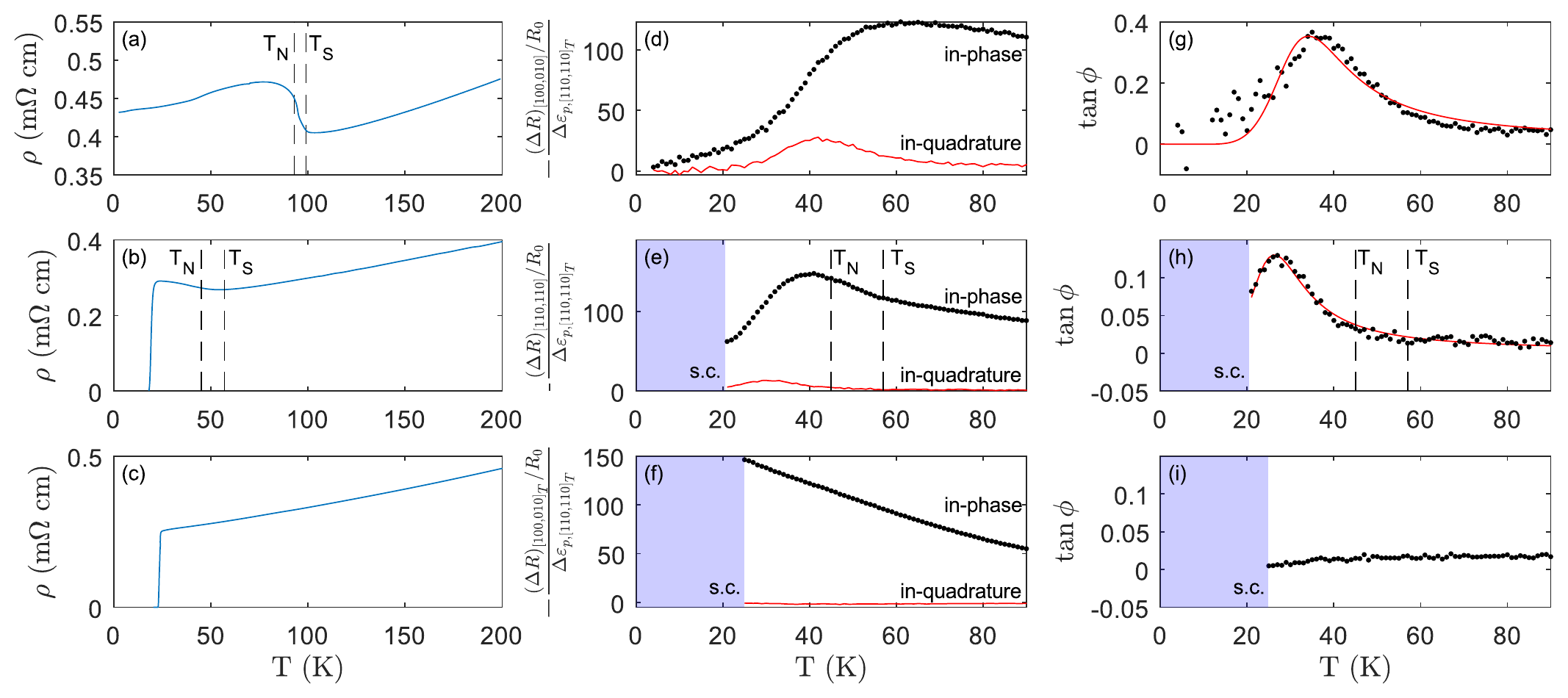}
    \caption{The resistivity of the three compositions of \BaFeCoAs are shown in (a-c) for samples of $x$ = 0.025, 0.052 and 0.07 respectively. 
    There are systematic uncertainties of order 40\% in the absolute scale for measurements of the resistivity, due to difficulty in measuring the thinness of the samples. 
    Clear signatures of the successive Ne\'el and structural transitions are evident in temperature derivatives of the resistivity, which we have marked on the corresponding plots.
    The resistance modulation of the three compositions of \BaFeCoAsNS, are respectively shown in (d-f) for an AC strain excitation of 105, 140 and 105 Hz respectively for a 10V excitation on the piezoelectric stack. 
    As the resistance change can be either in-phase or out of phase of the strain measured on the piezoelectric stack, we show both the in-phase component (black markers) and the in-quadrature component (solid red line), which we label in these panels.
    The ratio of the imaginary part to the real part of the resistance modulation is shown in the final column(g-i), with representative fits shown for the $x$ = 0.025 and 0.052 samples, as described in the text.
    Superconducting regions, where our resistance-based measurements are precluded, are indicated by the shaded region with the "s.c." label. }
    \label{fig:overviews}
\end{figure*}

Finally, we note that we can verify that the glue is not introducing artifacts to the phase response of the sample to the deformation of the piezoelectric stack.
In latter sections of this paper, we present measurements of a sample of Ba(Fe$_{0.93}$Co$_{0.07}$)$_{2}$As$_{2}$, which remains tetragonal at all temperatures and for which there is no temperature dependence to the phase between the sample resistance and the displacement of the piezoelectric stack (See \cref{fig:overviews}(f)).

\subsection{Response of a Material with a General Compliance }

The above treatment can be generalized for any sample with complex compliance $J_S(\omega)=J_1(\omega) - iJ_2(\omega)$ attached to the side of a piezoelectric stack via a glue with compliance $J_g(\omega)$.
As before, the stress on all components must be equal as in \cref{eq:sigmas}, and the total lengths of the various components are still given by \cref{eq:lengths}.
From these starting points, we find that the sample strain satisfies
\begin{equation}
    \varepsilon_s = J_s \sigma = \frac{J_s}{J_g}\left( \frac{L_p}{2L_{0g}}\varepsilon_p - \frac{L_s}{2L_{0g}}\varepsilon_s \right)
\end{equation}
or 
\begin{equation}
    \Gamma = \frac{\varepsilon_s}{\varepsilon_p} = \frac{L_p}{L_s} \frac{\tilde{J}_S}{1 + \tilde{J}_S }
\end{equation}
where $\frac{J_S}{J_g} \frac{L_s}{2L_{0g}} =\tilde{J}_s$ is the normalized compliance of the sample.
Then, the phase difference between the strain on the piezoelectric stack and the strain of the sample is
\begin{equation}
\tan\phi = 
\frac{\tilde{J}_2}{\tilde{J}_1 + \left|\tilde{J}_s\right|^2}
=\frac{J_2}{J_1}\frac{1}{1+ \frac{L_s}{2L_{0g}}\frac{\left|J_s\right|^2}{J_g J_1}}
,
\end{equation}
which can be used to fit various empirical forms of the compliance, including those proposed by Cole and Cole\cite{ColeCole}, Havriliak and Negami\cite{Havriliak1967} or by Nowick and Berry\cite{NowickBerryIBM}.


\begin{figure}
    \centering
    \includegraphics[width =\linewidth]{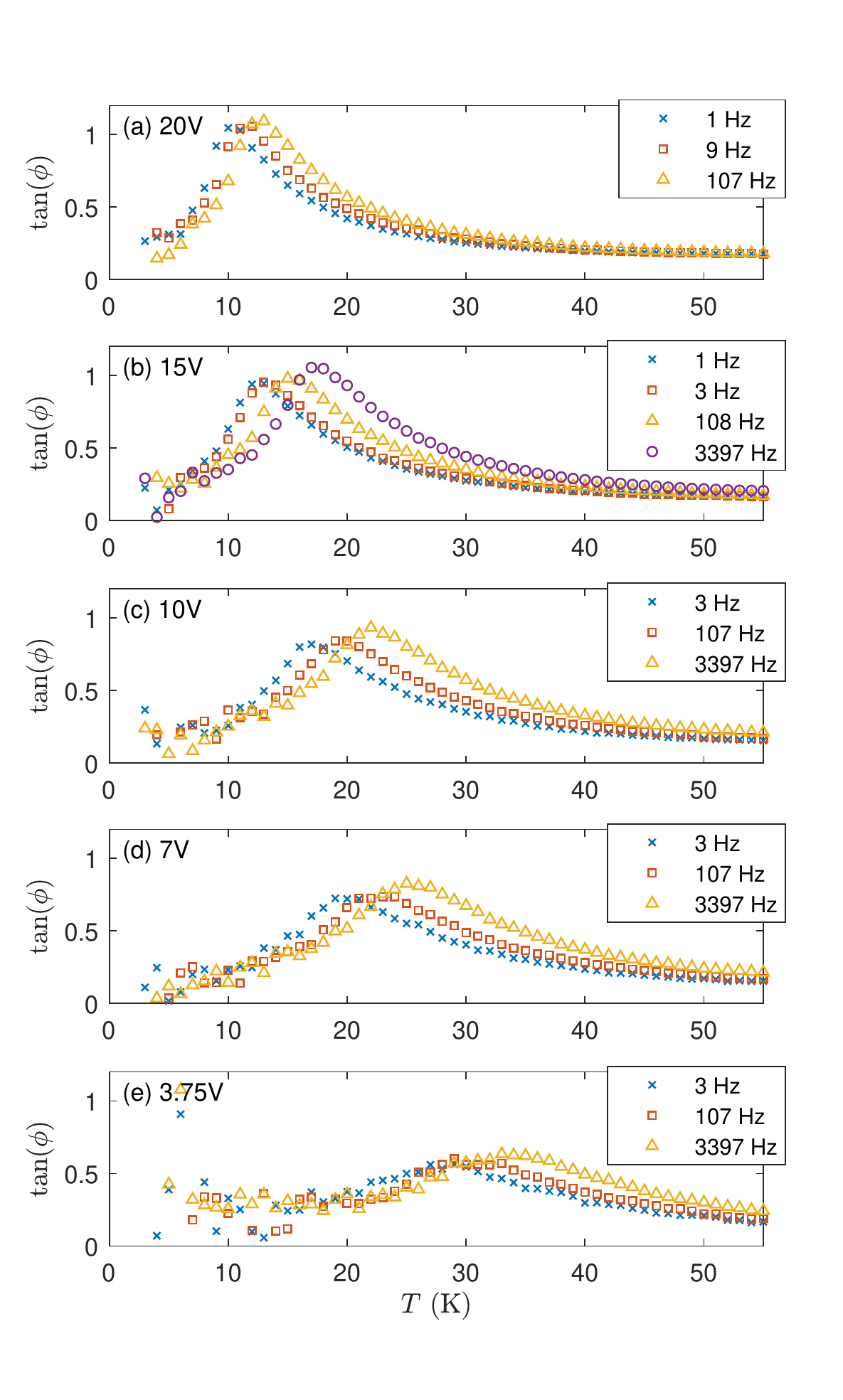}
    \caption{ 
    The phase lag $\phi$ of the resistance of a sample of Ba(Fe$_{0.975}$Co$_{0.025}$)$_{2}$As$_{2}$ relative to the expansion and contraction of the piezoelectric stack on which the sample is mounted.
    Each panel shows $\tan(\phi)$ for a set of traces with different frequencies and with a common amplitude of the voltage excitation supplied to the piezoelectric stack indicated in the upper left corner.
    }
    \label{fig:amplitudefrequencydependence}
\end{figure}

\section{Methods} 

Samples of \BaFeCoAs with compositions of $x=0.025$, $x=0.052$, and $x=0.070$ were grown from a self-flux method \cite{Chu_Determination_2009}.  
Samples were exfoliated to a thickness between 0.02 and 0.04 mm, then cut into a rectangular shape.
In-plane dimensions ranged from 0.8 mm to 2.0mm along the long axis of the crystal along which current flows, and approximately 0.45 mm to 0.8 mm in the perpendicular direction.  
Electrical contacts were made using ChipQuik SMD291AX10T5 solder.
Samples were adhered to the surface of a Piezomechanik ``PSt150/5x5/7 cryo 1'' using AngstromBond AB9110 LV epoxy, as shown in \cref{fig:schematic}(a).

The sample with $x=0.052$ was cut into a skinny bar with long axes oriented along the $[110]_T$ axis, contacted in a four-point measurement and affixed to the piezoelectric stack along the poling direction of the stack.
In this configuration, each domain is oriented with either the $a$ or $b$ axis along the direction of the current, and we measure the resistance along the $[110]_T$ axis, which we denote $R_{[110]_T,[110]_T}$.

Samples with $x=0.025$ and $x=0.07$ were cut with axes oriented along those of the tetragonal unit cell and contacts were placed in a five-point Hall-bar configuration; these samples were adhered to the piezoelectric stack with long axes of the sample at a 45 degree angle to the axes of the piezo stack.
This configuration is similar to the one used for the $x=0.052$ sample, in that the orthorhombic $a$ and $b$ axes of the piezoelectric stack are still oriented parallel to one of the axes of the piezo stack, thus the expansion and contraction of the piezoelectric stack can energetically favor one domain over another.
However, rather than measuring the change in a longitudinal resistance $R_{[110]_T,[110]_T}$ which has a large background resistance, this configuration uses the transverse voltage contacts of the hall bar, which measure $R_{[100]_T,[010]_T}$ as we have done previously\cite{Shapiro_Measurement_2016}.
Though this configuration may appear counter-intuitive, it is intuitive to observe that in isotropic materials that have been mechanically deformed so as to break the equivalence of the $[110]_T$ and $[1\bar{1}0]_T$ axes by an induced strain, then the transverse resistance satisfies 
\begin{equation}
R_{[100]_T,[010]_T} \propto (\rho_{[110]_T,[1\bar{1}0]_T}-\rho_{[1\bar{1}0]_T,[110]_T}).
\end{equation}
This highlights that the transverse voltage in this configuration is a measure of the symmetry breaking between the orthorhombic $a$ and $b$ axes, so that changes in the distribution of domains changes the apparent transverse voltage measured.
We shall show the phenomena we describe in this text can be robustly observed in the multi-domain phase in both configurations.

For both configurations, we source a current into the sample at frequency $\omega_c$ and measure the changes in a resistance $R$ caused by mechanical deformation at frequency $\omega_s$.
The combination of these two AC excitations on the sample produces an amplitude modulated voltage, which we demodulate using a Stanford Research 860 lock-in amplifier in dual-reference mode to obtain the deformation-induced change in resistance, $\Delta R$, which changes at the strain frequency $\omega_s$, as described previously\cite{HristovMeasurement2018}. 
In our analysis, this quantity is then divided by the unmodulated longitudinal resistance of the sample, which we denote $R_0$, following previous works of Chu, et al. \cite{chu2012divergent} and Tanatar et al.\cite{TanatarPRL}, which have shown that this quantity is a more appropriate measure of the electronic and structural anisotropy, and that this quantity better enables the comparison of measurements made at different temperatures.
The unmodulated resistance of the sample was also used as an in-situ thermometer to verify the absence of significant heating from the piezoelectric stacks, which arises from driving these stacks with large amplitude and high frequency excitations.

In these experiments, the range of mechanical strains induced in samples is small compared to the differences in lattice constants between domains. 
The piezoelectric apparatus we use lacks the dynamic range to make mono-domain samples, and its effects are therefore assumed to be weak and perturbative on the global population of domains.

\section{Results}

\Cref{fig:overviews} presents the resistivity (a-c) and the resistance oscillation resulting from AC stress(d-f) of three samples of \BaFeCoAs that span the orthorhombic region of the phase diagram. 
The sample resistances, or their temperature derivatives, are used to identify the Ne\'el and structural transitions shown in panels (a), (b), (e), and (h); this technique of identifying the phase transitions has previously been established in refs. \onlinecite{Chu_Determination_2009,Ni2008}.
Panels (d-f) show the normalized resistance modulation of the sample, $\Delta R/R_0$, in response to an excitation of 10V on the piezoelectric stack with a frequency of between 100 and 140 Hz. 

Clear signatures of an out of phase component of the resistance response to strain becomes evident in panels (d) and (e).
To our knowledge this is the first such observation of an out-of-phase response of resistance to an external stress. 
Importantly, the peak in the in-quadrature resistance response occurs well below Ne\'el and structural transitions.
For the $x=0.025$ sample, the transitions are at 94K and 99K, respectively, while the peak in the in-quadrature resistance response occurs at around 40K.
For the $x=0.052$ sample, the transitions temperatures are depressed to 45K and 57K, respectively, and the in-quardrature peak of the resistance response is suppressed to around 25~K. 
Finally, the sample with $x=0.07$ remains tetragonal at all temperatures and there is no observable peak in the in-quadrature response of the resistivity.
That the peak is only observed in the multi-domain state, and occurs below the transition temperature is qualitatively suggestive of an origin associated with domain wall motion.

The in-phase response of the resistance change to external stress also shows important qualitative indications that the resistivity response measured in this experiment comes from anelastic relaxation.
The peak in the in-quadrature response for $x=0.025$ and $x=0.052$ is coincident with a sharp decrease in the observed in-phase response shown in panels (d) and (e). 
This is typical of materials undergoing anelastic relaxation: decreasing temperature prevents thermally-activated relaxation of internal degrees of freedom, which in-turn decreases the sample compliance.
However, as measurements of resistance change due to mechanical modulation also depend on the value of the elastoresistivity coefficient, which may be temperature dependent and which have not previously been measured in this regime, we treat the in-phase and in-quadrature amplitudes merely as qualitative indicators.
Instead, we consider the relative phase between sample resistance and deformation of the piezoelectric stack, which cancels the temperature dependence of the elastoresistivity.

To make quantitative analysis of the AC change in resistance possible, we plot the ratio of the in-quadrature and in-phase resistance responses in panels (g) to (i) of \cref{fig:overviews}. 
We then fit these measurements according to our model for the anelastic relaxation of a standard linear solid model on the surface of a piezoelectric stack, \cref{eq:tanphi2}.
The best fits are shown in solid lines.
For comparison it is clear that no such peak occurs in the tetragonal sample with $x=0.07$ in panel (i).
 
 A systematic study of the amplitude and frequency dependence of $\tan\phi$ for the $x=0.025$ sample is shown in \cref{fig:amplitudefrequencydependence}.
 Each panel displays a set of traces taken with a common voltage excitation to the piezoelectric stack as a function of temperature.
 For each amplitude, the strain frequency was varied between 2 and 3.5 decades to produce the multiple traces on each panel. 
 At fixed amplitude, an increase in the frequency results in an increase in the peak position, as expected for anelastic processes.
 More surprisingly, there is a large dependence of $\tan\phi$ on the voltage supplied to the piezoelectric stack and therefore on the amplitude of stress induced in the sample.
 Increasing the amplitude of the mechanical perturbation on the sample dramatically decreases the temperature at which the peak in $\tan\phi$ is observed. 
 For example, the peak response to a 107 Hz mechanical excitation decreases from 31 K for a 3.75 V excitation to the piezoelectric stack to 13K for a 20V excitation.
The peak amplitude of $\tan\phi$ also increases with increasing mechanical excitation amplitude, to a peak value of approximately 1, similar to what has been found for other domain wall motion studies\cite{HarrisonSaljePhysRevB}.

\begin{figure}
    \centering
    \includegraphics[width =\linewidth]{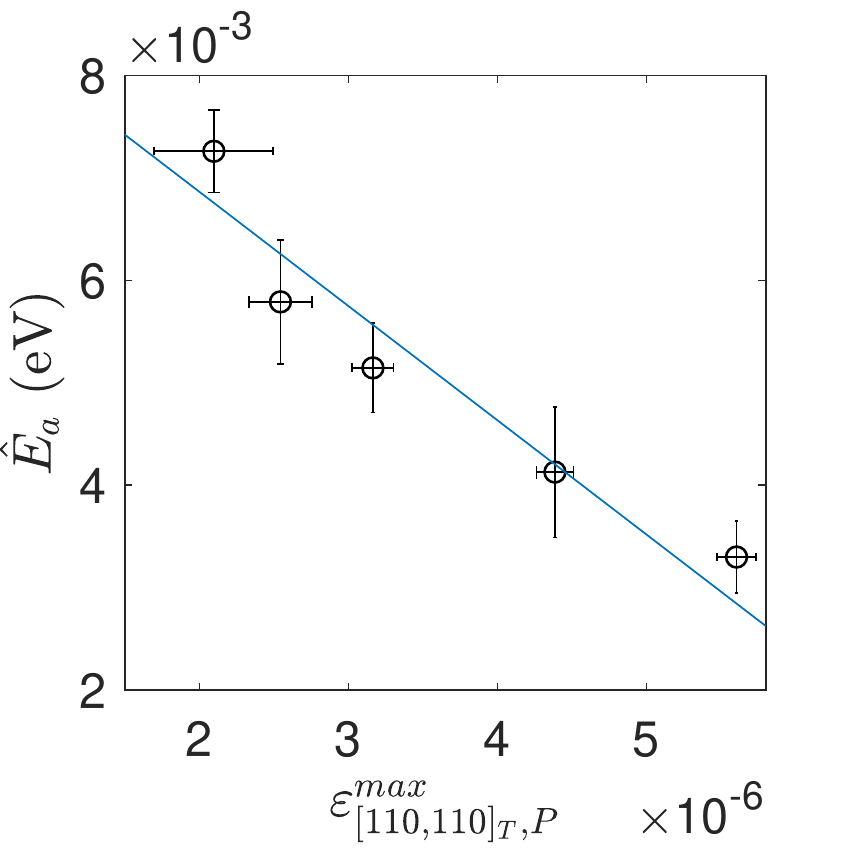}
    \caption{Estimate of the effective domain wall activation energies in  Ba(Fe$_{0.975}$Co$_{0.025}$)$_{2}$As$_{2}$ as a function of the strain amplitude, as measured by a strain gauge mounted to the piezoelectric stack parallel to the crystallographic $[110]_T$ axis.}
    \label{fig:activationvsamplitude}
\end{figure}

To extract the dependence of the anelastic relaxation process on the amplitude of the strain, we fit $\tan\phi$ in the region around each maximum.
\footnote{There is a small experimental offset to $\tan\phi$ at high temperatures, possibly due to imperfect synchronization of instrument electronics. As the fitted functional form in \cref{eq:tanphi2} tends to 0 in the limit of high and low temperatures, a restriction of the fitting window is necessary for convergence of the nonlinear fitting process.}
Around the temperatures $T_m$ at which there is a maximum in $\tan\phi$, we use a window of size 0.75$T_m$.
We fit all traces of a given panel to \cref{eq:tanphi2} and obtain a statistical estimate of the activation energy $E_a$.
Because the strain-per-volt of the piezoelectric is temperature dependent (though insensitive to strain frequency over the frequencies measured here), we also obtain the average and variance of the strain amplitude over that same measurement window.
The resulting estimates of domain wall activation energy are shown as function of strain amplitude in \cref{fig:activationvsamplitude}.

As a function of the strain amplitude, the estimated activation energy decreases from approximately 7.4 meV to 3.2 meV as the strain is increased from approximately 2ppm to over 6ppm. 
Due to variation in both the strain amplitude and the estimate of the activation energy, a weighted orthogonal least squares regression to a linear function to obtain estimates of $\frac{dE_a}{d\varepsilon^0_{[110,110]_T}} = -1115\pm 196$ eV, and $E_a(\varepsilon\to 0)=9.09 \pm 0.74 \times 10^{-3}$ eV.

\section{Discussion}

The dependence of the estimated activation energy on the amplitude of mechanical perturbation on the sample can be understood qualitatively from a simplified perspective. 
To begin, we distill the microscopic configuration of domain states into a collection of double-well energy potentials, each of the two states representing a different configuration of domain wall boundaries relative to defects inside the sample. 
Applying a stress to the sample, by exciting the piezoelectric stack in our apparatus, changes the energy difference between these nearby states and causes thermally activated relaxation from one domain configuration to another.
In particular, relative expansion along the $[110]_T$ axis increases the relative energy of orthorhombic domains with the shorter axis aligned to $[110]_T$ and decreases the energy of the other domain types.
In order for domain wall reconfiguration to occur, the domain wall boundary must pass through an intermediary state with an energy that is greater by an amount $E_a$.
As the amplitude of the stress is increased, at some point the energy difference between the two energy minima becomes comparable to the activation energy $E_a$, so that the relaxation rate during the AC stress cycle becomes much greater at maximum amplitude stress than the relaxation rate near zero stress.
In our technique, we measure only the $\omega_s$ component of the resistance change, and so are sensitive only to the cyclically averaged relaxation rate, though presumably signatures in higher harmonics could be investigated in the future.
Therefore, the relaxation rate increases with increasing strain amplitude, leading to a decrease in the activation energy apparent from fits of the Debye relaxation in \cref{eq:tanphi2}. 

At present, the energies extracted for the activation energy of the domain wall motion at vanishing strain from these measurements, $E_a(\varepsilon\to 0) =9.09 \pm 0.74 \times 10^{-3}$ eV, are 
lower than what is reported from resonant ultrasound ($E_a\in$ ($0.03$eV, $0.06$eV) ) by Carpenter, et al. \cite{Carpenter_2019} 
However, there are a number of experimental factors that could account for this discrepancy, which is about 4 times larger than what is found here from measurements of the resistivity response to mechanical perturbation.

First, this discrepancy in fitted activation energies can come from the effects of inhomogeneity, which serves to broaden the peak and reduce the apparent activation energy. 
In this scenario, the distribution of relaxation times can be parameterized as a log-normal distribution of width $\beta$, from which it would follow that $\hat{E}_a= \frac{E_a}{r_2(\beta)}$ where $r_2(\beta)$ is a monotonically increasing function of $\beta$ with $r_2(0)= 1$, as described in Ref. \onlinecite{NowickBerryIBM}.
This effect can also be present in resonant ultrasound, and indeed reports of this technique have stated a difficulty in fitting both the peak position and peak width simultaneously.  

A second important effect comes from the large strains induced in our measurements.
These are larger than what is used for resonant ultrasound because we use higher voltages to excite the piezoelectric stack, and we also mount the sample to the piezoelectric stack instead of mounting a long sample between thin piezoelectric transducers, which produces a mechanical disadvantage and reduces the amplitude of strain in the sample.
An inspection of \cref{fig:activationvsamplitude} suggests that if the relation between activation energy and mechanical perturbation is non-linear, then the zero strain limit $E_a(\varepsilon\to 0)$ could be considerably greater, depending on the precise functional form.

\section{Conclusion}

We present measurements of the resistance response of \BaFeCoAs to an external mechanical stress in the multi-domain state. 
Because the amplitude of resistance change to mechanical perturbation depends on the elastoresistivity coefficient, which has not been isolated in this region, we choose instead to focus on the phase lag between the AC resistance modulation and the stress on the sample.
We developed a model which explains in simple terms how the phase lag $\phi$ between the resistance, which is a function of the domain configuration in the sample, and the stress on the sample depends on the complex compliance of the sample.

Well below the structural and Ne\'el transitions, the resistance response to mechanical deformation in both $x=0.025$ and $x=0.052$ has a peak in the in-quadrature response.
Measurement for an $x=0.07$ sample, which remains tetragonal at all temperatures, shows no such peak, which excludes the possibility that the measurements in $x=0.025$ and $x=0.052$ are merely artifacts from sample preparation.
Instead, we proceed to demonstrate that the temperature, amplitude, and frequency dependence of $\tan(\phi)$ is qualitatively and quantitatively consistent with anelastic relaxation of domain walls with an activation energy of approximately $E_a = 9.09 \pm 0.74 \times 10^{-3}$ eV in the zero strain limit, which is on a similar energy scale to previous measurements attributed to polaronic defects\cite{Carpenter_2019}.
As the amplitude of mechanical perturbation is increased, we find $\frac{dE_a}{d\varepsilon^0_{[110,110]_T}} = -1115\pm 196$ eV.

\begin{acknowledgements}
This work was supported by the Department of Energy, Office of Basic Energy Sciences, under Contract No. DE-AC02-76SF00515. M.S.I. was partially supported by the Gordon and Betty Moore Foundations EPiQS Initiative through Grant No. GBMF4414. A.T.H and J.C.P were supported by a NSF Graduate Research Fellowship through Grant No. DGE-114747. J.C.P. also acknowledges funding from a Gabilan Stanford Graduate Fellowship and a Lieberman Fellowship.
\end{acknowledgements}

\bibliography{refs}

\newpage

\appendix
    \section{Offset Strain Dependence of Anelastic Response}
\begin{figure}[b]
        \centering
        \includegraphics[width=\linewidth]{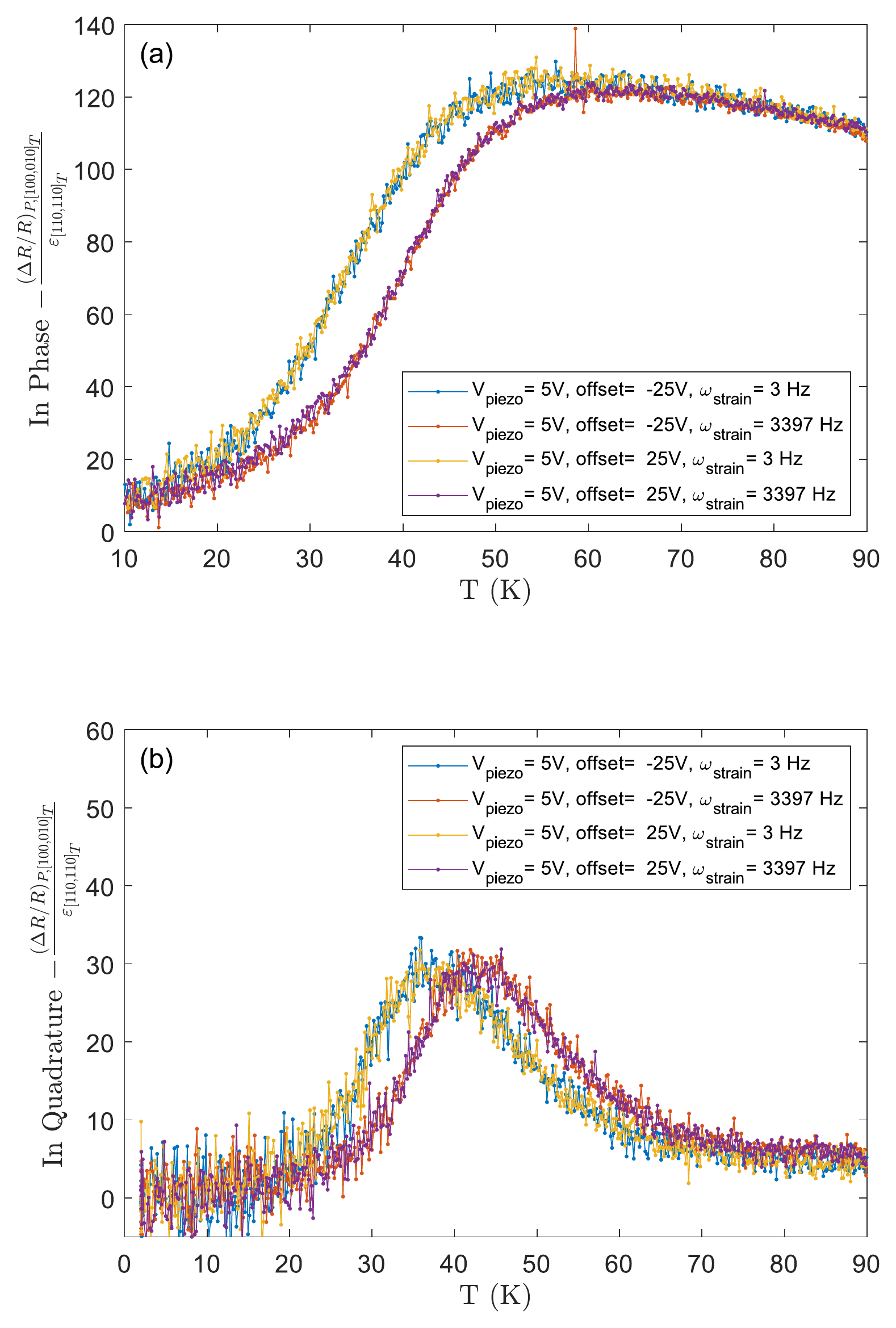}
        \caption{The resistance response of \BaFeCoAs (x=0.025) to a 5~V AC excitation on the piezoelectric stack is shown for DC offsets of $\pm 25$V and AC excitation frequencies of 3~Hz and 3.4~kHz. Panel (a) shows the in-phase response of the resistance to strain, while panel (b) shows the in-quadrature response.}
        \label{fig:appoffset}
    \end{figure}

    To test the linearity of the sample resistance response to strain, and to ensure the mechanical integrity of the samples used in this study, we applied varying DC offsets to the AC excitation on the piezoelectric stack for our $x=0.025$ substituted sample of \BaFeCoAsNS. Representative traces taken at $\pm$25V offset on the piezoelectric stack are shown in \cref{fig:appoffset}.
    The two traces at 3~Hz are clearly distinguishable from the traces at 3.4~kHz and the traces at the same frequency are indistinguishable from one another.
    This clearly indicates that the responses measured in this sample are not sensitive to the range of offset strains used in this measurement.
    While it is not possible to cancel the offset strain from the differential thermal expansion of the piezoelectric stack and the sample, these measurements provide provide no evidence such a strain is influencing measurements.
    Instead, these measurements appear to confirm that the mechanical deformations used in this study are not sufficient enough to de-twin the samples, and instead provide small perturbations to the global domain population.
    
    \section{Very Low Frequency Strain Response}
\begin{figure*}
        \centering
        \includegraphics[width= 0.8\linewidth]{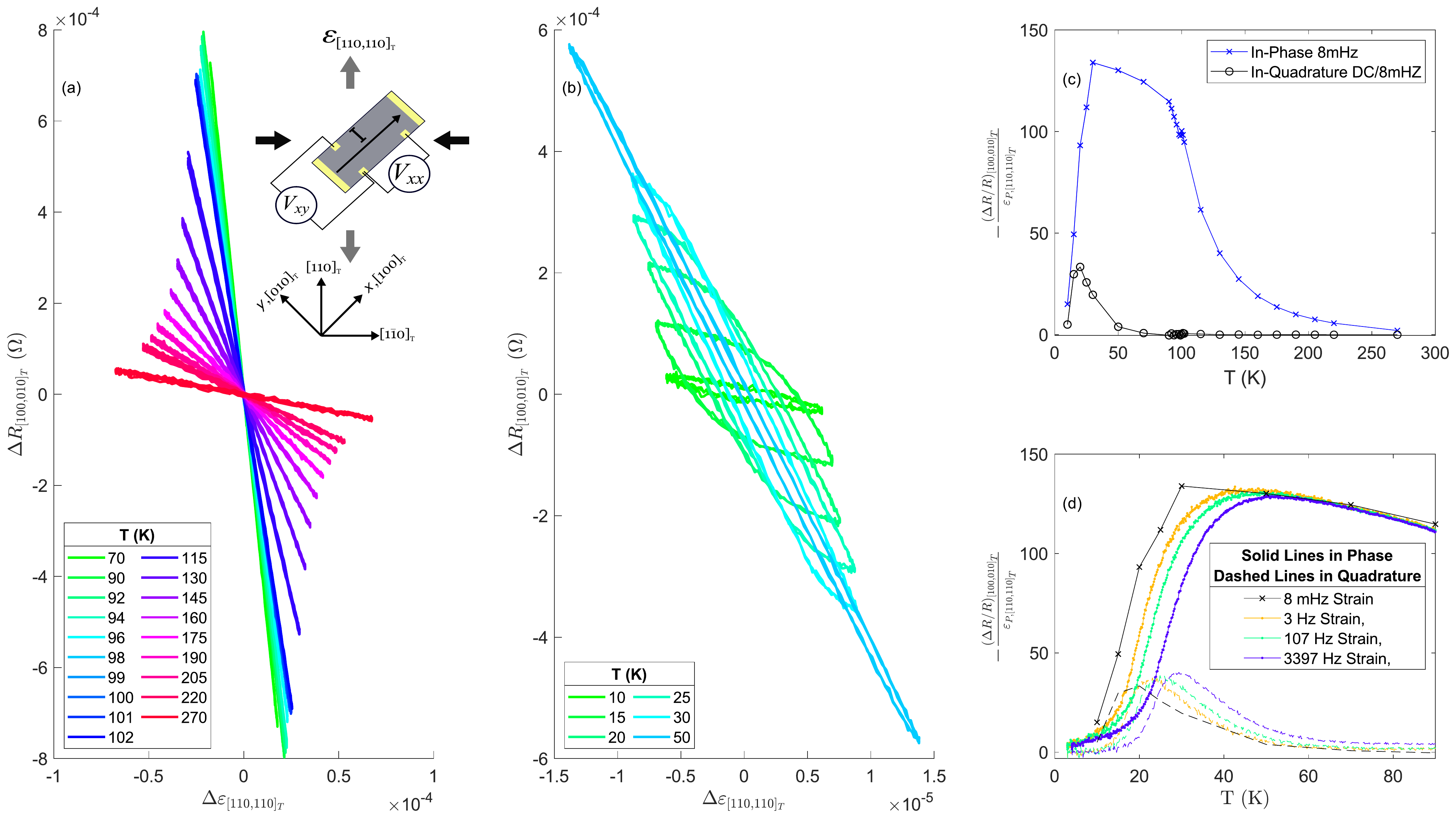}
        \caption{(a) The change in transverse resistance of a sample of \BaFeCoAs ($x=0.025$)  against the equivalent change from the restive strain gauge mounted on to the pizeoelectric stack, for a range of temperatures indicated in the legend.
        The inset shows the sample (grey rectangle) with electrical contacts (yellow) aligned at a 45 degree angle to the piezoelectric axes (grey arrows) which elongate along the $[110]_T$ crystallographic axis and contract along the perpendicular axes.
        (b) Below 50K, hysteretic behavior manifests in the resistance response of the sample relative to the resistance of the strain gauge.
        (c) The in-plane and in-quadrature resistance response of the sample with respect to the deformation of the piezoelectric stack, as obtained from fitting both the strain gauge and sample resistances to sinusoidal functions.
        (d) The resistance response to the 8~mHz mechanical perturbation provided by a 10~V sine wave to the piezoelectric stack is plotted alongside measurements taken using an AC mechanical modulation technique for frequencies ranging from 3~Hz to 3.4~kHz using the same excitation amplitude. 
        }
        \label{fig:DC}
    \end{figure*}

    The resistance response to a quasi-static deformation of the piezoelectric drive apparatus is shown in \cref{fig:DC}.
    In this procedure, the voltage on the piezoelectric stack is incrementally stepped every 5 seconds, with the sequence of voltages $V_n$ tracking $V_n = 10\sin(2\pi n/160)$~V, which comprises a primary strain frequency ($\omega_s = 2\pi /(160\times5)$ Hz) of approximately 8~mHz, with a secondary contributions from higher-harmonics from the step-wise voltage changes.
    The sample and strain gauge resistance is measured at the end of the five seconds by a Stanford Research 860 or 830 lock-in amplifier with a time constant, 100ms, such that the instrument settling time is well below five seconds.
    
    At high temperatures, the resistance response to mechanical deformation is linear and clear, providing evidence of the mechanical integrity of the sample, as shown in panel (a) of \cref{fig:DC}.
    As with other measurements in this work, a hysteresis becomes apparent at temperatures below 50~K. To quantify this hysteresis, the normalized resistance of the sample and the strain measured by the strain gauge are fit functional forms to 
    \begin{equation}
        \left(\frac{\Delta R}{R}\right)_{[100,010]_T} = A\sin\left(\frac{2\pi n}{160}\right)+B\cos\left(\frac{2\pi n}{160}\right)
    \end{equation}
    \begin{equation}
        (\varepsilon)_{[110,110]_T} =  C\sin\left(\frac{2\pi n}{160}\right)+D\cos\left(\frac{2\pi n}{160}\right)
    \end{equation}
    at each temperature. For that temperature, the in-phase and in-quadrature components of the resistance response to strain are obtained from the real and imaginary parts of 
    \begin{equation}
        \frac{\left(\frac{\Delta R}{R}\right)_{[100,010]_T} }{\varepsilon_{[110,110]_T}} =\frac{A+iB}{C+iD}
    \end{equation}
    which is shown in panel (c) of \cref{fig:DC}.
    Comparing this response to other measurements for an identical amplitude AC voltage driving the piezoelectric device, we find from panel (d) that this ultra-low frequency response is broadly consistent with the phenomenology of higher frequency measurements, and the observed phenomena likely have a common origin.
    A precise fit of the temperature dependence of $\tan\phi$ measured by this low-frequency strain technique is precluded by the long measurement times necessary to perform measurements using this quasi-static technique at low temperatures, where the strain response of the piezoelectric apparatus is diminished. 
    As a result, three such measurements performed below 20K which is sufficient for the qualitative comparison to measurements performed with higher frequency mechanical perturbations, as shown in (d) of \cref{fig:DC}.
    
\end{document}